\documentclass[referee,a4paper,12pt,traditabstract]{jswsc} 
\usepackage{graphicx}
\usepackage{subfigure}
\usepackage{epstopdf}
\usepackage[mathlines]{lineno}
\usepackage[authoryear,round]{natbib}
\usepackage[backref]{hyperref}
\usepackage{newtxmath}
\usepackage{mathtools}
\usepackage{url}
\bibpunct[, ]{(}{)}{;}{a}{,}{,}

\hypersetup{colorlinks=true,citecolor=cyan,urlcolor=cyan,linkcolor=blue} 

\begin{document}

\title{Multi-scale Image Preprocessing and Feature Tracking for Remote CME Characterization}

\titlerunning{Feature Tracking for Remote CME Characterization}

\authorrunning{Stepanyuk, Kozarev, Nedal}

\author{O. Stepanyuk
      \inst{1}
      \and
      K. Kozarev\inst{1}
      \and
      M. Nedal\inst{1}
      }

\institute{
          Institute of Astronomy and National Astronomical Observatory\\
          Bulgarian Academy of Sciences\\
          Sofia, Bulgaria\\
          \email{\href{mailto:ostepanyuk@astro.bas.bg}{ostepanyuk@astro.bas.bg}}
          }

\abstract{Coronal Mass Ejections (CMEs) influence the interplanetary environment over vast distances in the solar system by injecting huge clouds of fast solar plasma and energetic particles (SEPs). A number of fundamental questions remain about how SEPs are produced, but current understanding points to CME-driven shocks and compressions in the solar corona. At the same time, unprecedented remote and in situ (Parker Solar Probe, Solar Orbiter) solar observations are becoming available to constrain existing theories. Here we present a general method for recognition and tracking on solar images of objects such as CME shock waves and filaments. The calculation scheme is based on a multi-scale data representation concept \`a trous wavelet transform, and a set of image filtering techniques. We showcase its performance on a small set of CME-related phenomena observed with the SDO/AIA telescope. With the data represented hierarchically on different decomposition and intensity levels, our method allows to extract certain objects and their masks from the imaging observations, in order to track their evolution in time. The method presented here is general and applicable to detecting and tracking various solar and heliospheric phenomena in imaging observations. It holds potential to prepare large training data sets for deep learning. We implemented this method into a freely available Python library.}

\maketitle

\section{Introduction}
\label{s1}
Solar eruptive events are complex phenomena, which most often include solar flares, filament eruptions, coronal mass ejections (CMEs), and CME-driven shock waves. CME-driven shocks in the corona and interplanetary (IP) space are considered the main producer of solar energetic particles (SEPs), primarily via the diffusive shock acceleration (DSA) and shock drift acceleration (SDA) processes \citep{Reames:2021}. Considerable effort has been expended in characterizing and modeling the various aspects of SEP acceleration, under idealized conditions \citep{Vainio:2008, Sokolov:2009, Kozarev:2013}. Recent efforts have focused on building more realistic models for CME-driven shocks and their acceleration of SEPs, directly driven by real data. \citep{Vourlidas:2012, Kwon:2014, Kozarev:2015, Kozarev:2019}.

Knowledge of how CME-driven shocks interact with the three-dimensional coronal magnetic fields and plasma is crucial for understanding how efficiently they accelerate particles, as well as how much SEP fluxes may spread in heliospheric longitude and latitude \citep{Rouillard:2016}. A specific requirement for modeling SEP acceleration by DSA is deducing the shape of the coronal shock fronts as accurately as possible from the observations. This is because the local orientation of the magnetic field to the shock front strongly modulates the particle acceleration process \citep{Guo:2013}. Furthermore, it has been shown that often in the early stages of eruptions the CME over-expands in the lateral direction, changing the overall shape of the compressive front ahead of it \citep{Bein:2011, Temmer:2016}. Thus, it is imperative to move beyond the idealized descriptions of the shock surfaces regularly used to model their evolution \citep{Vourlidas:2012, Kwon:2014, Rouillard:2016}, and to employ more advanced techniques.

Characterizing coronal shocks is possible using remote measurements from instruments on modern heliospheric space missions. Extreme ultraviolet (EUV) imaging with telescopes such as the EUV Imagers \citep[EUVI]{Wuelser:2004} on the STEREO mission and the Atmospheric Imaging Assembly \citep[AIA]{Lemen:2012} instrument on the Solar Dynamics Observatory (SDO) has proven particularly useful for detecting and characterizing large-scale shocks, also known as EUV waves or coronal bright fronts \citep[CBFs]{Long:2011}, because of its unprecedented temporal and spatial resolution, and multi-wavelength coverage. Such time-dependent characterization can further be used to inform models of energetic particle acceleration in the solar corona and inner heliosphere \citep{Kozarev:2016, Kozarev:2017, Kozarev:2019}. Development of automated solar feature-detection and identification methods has increased in recent years due to the demand from Big Data. An overview of fundamental image processing techniques used in these algorithms is presented in \citet{Aschwanden:2010}. These techniques are used to detect many features in various types of observations at different heights in the solar atmosphere \citep{Perez-Suarez:2011}. Some previous work has focused on detecting sunspot groups and active regions in photospheric continuum images, magnetograms, and EUV images \citep{Curto:2008}.

The issue with most algorithms currently used in solar feature detection and tracking is the complexity of their processing chains, which makes them difficult to implement; that they are usable for very specific tasks, and that often they are applicable only to data from a single instrument. 
Specifically, the task of EUV wave recognition and tracking is complicated by their much weaker intensity compared with most other solar features, next to which they propagate and project. A number of algorithms for EUV wave detection of varying complexity exist \citep{Podladchikova:2005, Verbeeck:2014, Long:2014, Ireland:2019}. Of these perhaps the most advanced are the CorPITA \citep{Long:2014} and AWARE \citep{Ireland:2019}. The CorPITA algorithm employs percentage base difference images to fit multi-Gaussian shapes to flare-source-centered, sector-averaged portions on the wave along the solar disk. The AWARE algorithm uses more advanced pre-processing in persistence running difference images, to characterize the wave shapes along similar flare-centered sectors, and a random sample consensus (RANSAC) algorithm to select the features. However, using running difference images introduces spurious features and is not recommended for discovering the true (projected) wave shape \citep{Long:2011}. In addition, the persistence imaging approach tends to amplify the noise in the output, requiring additional post-processing (median filtering and closing operations). While both these algorithms are well automated based on flare onset signal, they have so far been applied only on the solar disk. What is more, their procedures are focused specifically on EUV waves.

In contrast, here we present Wavetrack - a generalized yet flexible framework for solar feature detection, implemented as an object-oriented, easily to integrate, and freely-available Python library\footnote{Hosted at \url{https://gitlab.com/iahelio/mosaiics/wavetrack}.}. In its core is a general method for recognition and tracking of features on solar images, such as CME shock waves and filaments. The different applications may be easily controlled by modifying a few threshold parameters. The calculation scheme is based on a multi-scale data representation concept \citep{Starck:2002}, combined with an \`a trous wavelet transform \citep{Akansu:1991, Holschneider:1989}, and a set of image filtering techniques.

In recent years Machine Learning methods have become more frequently applied in solar physics.  For instance, \citet{Szenicer:2019} used a combined CNN to produce the EUV irradiance map from SDO/AIA images, \citet{Li:2013} used a multi-layer model o predict solar flares based on sequential sunspot data, \citet{Kim:2019} applied generative adversarial networks (GAN) to generate the magnetic flux distribution of the Sun from SDO/AIA image. Nevertheless, usage of data-driven approaches for tracking of CME-related phenomena is currently limited due to insufficiency of training sets, and having a tool that would make preparation of such training sets a more easy task is one of the motivations behind current research. Wavetrack results can be easily converted into annotated training sets for teaching learning models to recognise various solar features in remote observations.

The paper is organized as follows: We describe the method of Wavetrack in Section \ref{s2}, followed by an application to several events and their analysis, presented in section \ref{s3}. We provide conclusions and a summary in Section \ref{s4}.

\section{Methodology}
\label{s2}

In this section, we discuss the overall methodology of the Wavetrack framework, and the various steps taken in processing the input data. We defer the most technical parts of the discussion to the Appendices of the paper.

\subsection{\`A trous Wavelet Decomposition}
\label{wavelet_decomposition}
In our study, we applied an \`a trous wavelet decomposition method, in which the data is passed through high and low-pass filters without decimation.This algorithm is more famously known as "algorithme a trous", which refers to inserting zeros in the filters. It was introduced by \citet{Holschneider:1989} and we used the approach similar to the one described in \citet{Starck:2002, Chui:1992, Stenborg:2003}. In previous observational studies, it has shown great capability for enhancing solar EUV and white light images, also improving the clarity and intensity of faint features, such as EUV waves \citep{Stenborg:2003, Stenborg:2008}.

Briefly, the algorithm works as follows. At each step of the algorithm, the image is convolved with the J-th iteration of the \`a trous wavelet kernel, which in its initial form $k_{0}$ has a form of a 5x5 matrix:
$$ k_{0} = 
\begin{array}{|ccccc|}
1 / 256 & 1 / 64 & 3 / 128 & 1 / 64 & 1 / 256 \\
1 / 64 & 1 / 16 & 3 / 32 & 1 / 16 & 1 / 64 \\
3 / 128 & 3 / 32 & 9 / 64 & 3 / 32 & 3 / 128 \\
1 / 64 & 1 / 16 & 3 / 32 & 1 / 16 & 1 / 64 \\
1 / 256 & 1 / 64 & 3 / 128 & 1 / 64 & 1 / 256 \\
\end{array}
$$
(see Appendix A, Eq. \ref{eq:a12} and Eq. \ref{eq:a13}).
The result is a wavelet coefficient matrix $c_j$. The wavelet scale for decomposition level $j$ is defined as $\omega_{j+1}=c_{j+1}-c_{j}$. The full \`a trous wavelet calculation scheme for a pre-selected set of $j=[1,J]$ decomposition levels consists of the following steps:\\
1) Set the wavelet coefficient $c_{0}$ equal to the original image; \\
2) Calculate each consecutive wavelet coefficient by convolving the previous wavelet coefficient with the corresponding iteration of the \`a trous wavelet kernel: $c_{j+1}=c_{j}*k_j$. $k_j$ is calculated by dilation of $k_0$ with a rate of $2^j$;\\
3) Calculate the scale of the discrete \`a trous wavelet transform as the difference between consecutive wavelet coefficients;\\
4) Normalize scales and coefficients for further analysis.\\

\begin{figure}[htp]
    \centering
    \includegraphics[width=12cm]{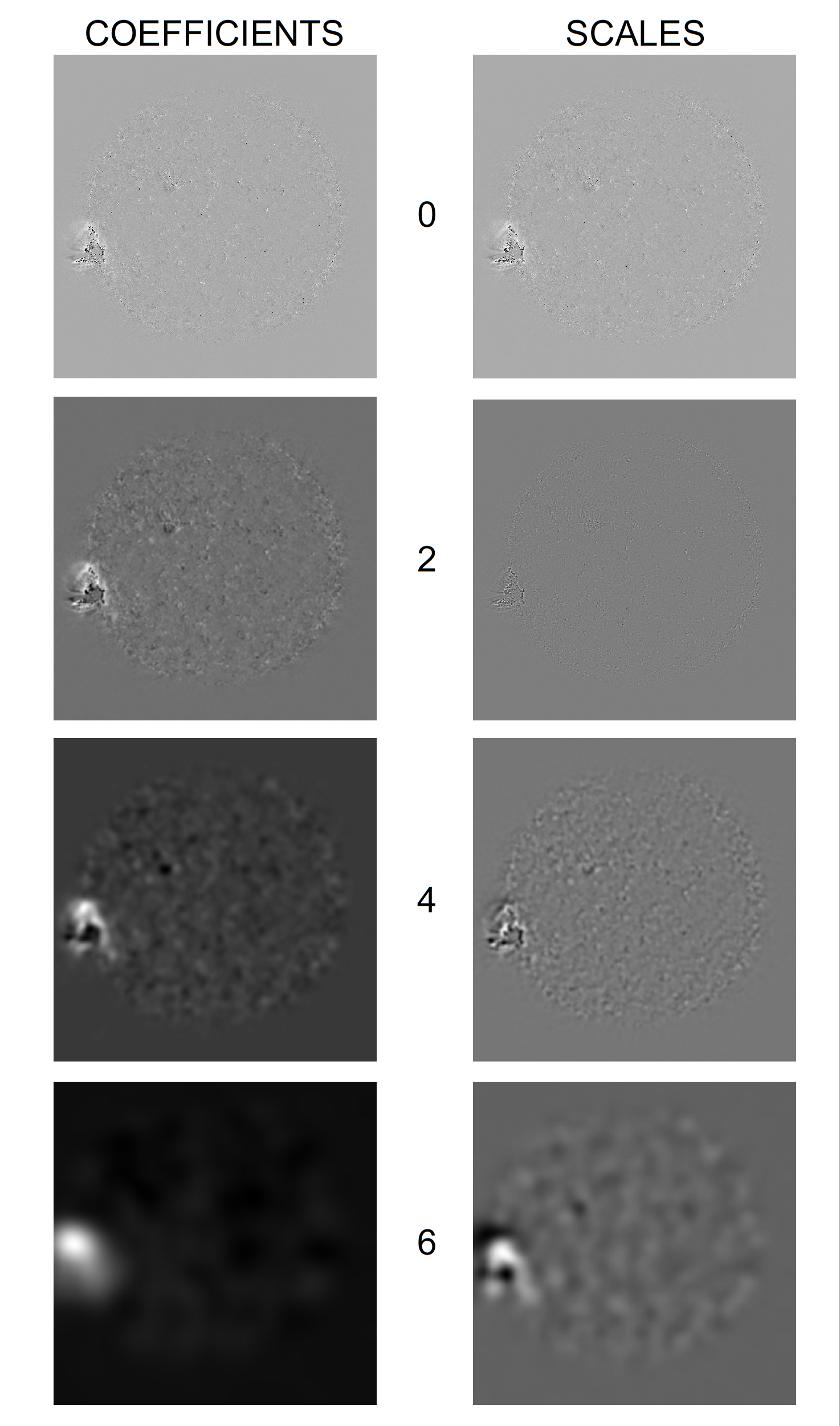}
    \caption{The figure shows wavelet coefficients and scales obtained by decomposing a normalized initial image (in this case, an AIA 193-channel base difference) after thresholding and shifting, so that the values are fit into [0,1] interval. Left panels: Wavelet coefficients at decomposition levels 0, 2, 4, 6. Right panels: Corresponding wavelet scales at the same decomposition levels.}
    \label{fig:a_trous}
\end{figure}

We have found the \`a trous wavelet decomposition to be the most efficient method when it comes to splitting solar images into a series, with each of the space-detail levels highlighted separately. Still, as Wavetrack was developed as a general framework for image recognition and tracking in solar physics, any other similar wavelet technique can be utilized, and the basic $J_{0}$ kernel can be replaced. Every j-th iteration of the kernel is initialized with zeros; positions in the wavelet scale matrix are then filled according to the values of the mother-wavelet $J_{0}$ kernel matrix, as multiple operations with diluting matrices are less computationally efficient.

\subsection{Structures. Object Criteria}
\label{sect_structures_object_criteria}
\begin{figure}[htp]
    \centering
    \includegraphics[width=12cm]{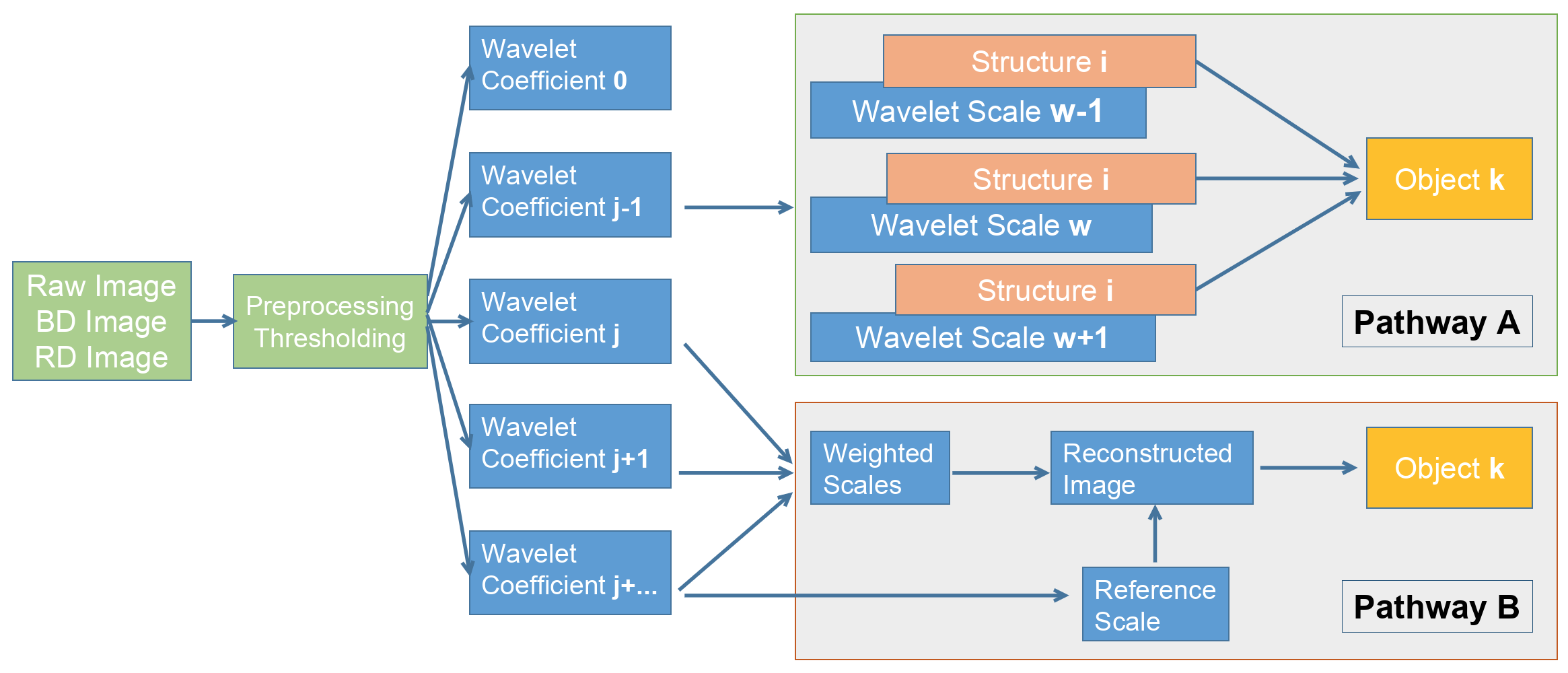}
    \caption{Image processing stages with Wavetrack software. Base Difference, Running Difference or RAW image is used as an input. After pre-processing, the image is decomposed with a-trous wavelet transform, and objects are labeled depending on the wavelet scales values}
    \label{fig:fig_scheme}
\end{figure}

By structure we mean a connected set of pixels on a single wavelet scale (See "Pathway a" from Figure \ref{fig:fig_scheme}).The Structure $S$ defines an Object on a scale $j$ if:   
\begin{equation}\label{eq:6}
\omega_{j}^{s}>\omega_{j-1}^{s}
\end{equation}
and
\begin{equation}\label{eq:7}
\omega_{j-1}^{s}>\omega_{j+1}^{s}
\end{equation}
where $\omega_{j}^{s}$ is the maximum value of the wavelet coefficient for the structure $S$. If the coefficient does not belong to any structure on the $j-1$ scale, then
\begin{equation}\label{eq:9}
\omega_{j-1}^{s}=0
\end{equation}
Otherwise it holds the maximum value for the structure:
\begin{equation}\label{eq:10}
\omega_{j+1}^{s}=\max \left\{\omega_{j+1, x_{1}, y_{1}}^{s}, \ldots, \omega_{j+1, x_{n} y_{n}}^{s}\right\}
\end{equation}
where all wavelet coefficients on a scale $j$ belong to the structure $S$ on that scale:
\begin{equation}\label{eq:11}
\omega_{j, x, y} \in S_{j}
\end{equation}


\subsection{The Wavetrack software. Concepts and processing routine}

The Wavetrack framework performs wavelet decomposition of observational data and automated feature recognition. Intensity thresholding, image posterization, median filtering, image segmentation and intensity levels splitting methods are available. These are defined below. They can be applied to each of the wavelet decomposition levels separately with different set of parameters. The image can be recomposed from weighted wavelet scales after application of filtering techniques.  The framework follows a modular approach and is built as a set of classes. Within a typical event processing routine every class represents one or a few image processing techniques, so that these classes act as building blocks allowing various decomposition and processing configurations and setups, depending on the input image characteristics and type of objects of interest.

\subsubsection{Input Data}
In this paper we use the term coronal bright front \citep{Kozarev:2015} to refer to the phenomenon that can be described as a shock wave, shock front, or bright front, defined as the anti-sunward edge of the shock sheath.
According to numerous previous studies \citep{Long:2011,Kozarev:2015}, the best SDO/AIA channels for observing CBFs are the ones centered on 193\AA~and 211\AA~wavelengths. In this work, we use 193\AA~data. The data is initially obtained via a standard AIA pipeline using SunPy package \citep{Sunpy:2020}.

Base images are created by averaging series of images a few minutes prior to the start of the eruption. Each time-step we start with base difference images which are constructed by subtracting base images from the current RAW image of the sequence. That allows to enhance the change in intensity, caused by CBF, omit static details and reduce noise. Running difference or persistence running difference images \citep{Ireland:2019} are used when necessary to determine the moment when the eruption starts.

Unlike the bright active regions, the CBFs observed in EUV are usually very dim features, which exist in a very narrow part of the extremely wide dynamic range of the raw image data. This fact generally makes these objects harder to recognize and track with standard techniques, particularly on relatively noisy AIA images. To overcome this obstacle, we pre-select the part of the dynamic range where the shock wave is revealed from the base difference image of the sequence.

Wavetrack can also be used to track filaments, and we present some examples further in the article. As they are formed with relatively cool and dense gas suspended above the surface of the Sun, held by magnetic fields, from the point of image processing they are generally less blurry objects with more distinct contours than CBFs; thus, they exist in lower parts of the image dynamic range. So within the typical Wavetrack setup used to track filaments, we use inverted AIA 193\AA~images and less posterization (defined below in Section \ref{multiscale_image_filtering}) on coefficients with higher bit rates ($>$8 bits, see below). Depending on the source data and velocity of the object of interest, Base Difference, Running Difference or Raw Intensity data may used for filaments. Wavetrack can be also used to create AWARE-like \citep{Ireland:2019} persistence running difference images.

Obtaining on-disk parts of CBFs and filaments appears to be a more difficult task than for the the off-limb parts. Regardless of the class of the event and type of difference images used as an input, on-disk parts of the object usually are surrounded by many low-scale details and noisy background with high pixel intensities. That leads to significantly different statistical distributions of pixel intensities, requiring different processing and decomposition setups, in base and running difference, as well as original images. For that reason these two parts of the image often are processed separately and the results are merged.

\subsubsection{Multiscale Image Filtering}
\label{multiscale_image_filtering}
Conceptually, iterative application of the \`a trous wavelet transform means decomposition of an image into a few separate levels of detail. Thus, processing of decomposition scales separately appears practical and gives advantages when the shape and size of the objects can be guessed. For any high-bit depth, low-contrast images, segmentation of objects becomes a non-trivial task due to the fact that it is unclear how to define object criteria unambiguously for the group of pixels. Image bit depth determines a number of intensity levels to which an image could be theoretically separated. Posterization is the conversion of a continuous gradation of tone to several regions of fewer tones and happens naturally when image bit depth has been decreased.

It could be assumed that the statistical dispersion of the pixel intensities for the structure (such as a shock wave or the filament) would be lower than for the noise or other surrounding signal, particularly for the disk. In other words, the structure is more "homogeneous" than its surroundings. If a wavelet coefficient reveals the object (and to a certain extent already skips the noise and irrelevant low-sized details), reducing bit depth may help further increase visual contrast and reveal shapes and contours. In this way, a gradient-contour operator applied on a reconstructed image (See Appendix \ref{image_gradient_appendix}) does not produce spurious contours. We empirically found that reducing wavelet coefficient resolution to 5 bits works best for dim and blurry objects; consequently, this bit depth reduction was used for all events from the current study, where we aimed to locate a CBF on or near the limb.\\

The scheme on Figure \ref{fig:fig_scheme} gives a general overview of the image processing stages with the Wavetrack software. For the AIA instrument data, the processing sequence for a single timestep goes as follows: \\
1) The Sunpy package is used to process the input data to AIA 1.5 level; \\
2) A Base Difference/Running Difference image is obtained; \\ 
3) Certain part of the dynamic range is selected by cutting off the data above and below high and low threshold values from the input parameters; \\ 
4) The image is decomposed with the \`a trous wavelet transform; \\
5) Relative thresholding (sigma units) is applied to each of the wavelet coefficients separately. The pixel intensities are assumed to follow a Gaussian distribution; \\
6) Bit depth of the wavelet coefficients is reduced; \\
7) Object masks are obtained via Pathway A or Pathway B (see Fig. \ref{fig:fig_scheme}). \\ 
 7.1) In case of large-scale off-limb objects (coronal shock waves), a simplified processing scheme is usually enough. The object mask is obtained by direct reconstruction of the image from wavelet scales with certain weight coefficients, and one of the higher level scales is applied as a "reference mask". \\
 7.2) In more complex image processing cases, such as, for instance, an early-stage on-disk filament, complete utilization of multi-scale image decomposition approach would be necessary. Here, wavelet scale values serve as object criteria (Described in Section \ref{wavelet_decomposition}). \\
8) Calculate gradient field of the reconstructed image; \\
9) Stand-alone masks of every object are obtained via segmentation routine (See Appendix \ref{image_segmentation_appendix}); \\
10) Calculate center of mass and geometrical center of the object (See Section \ref{centers_of_mass});\\
11) Estimate the object velocity based on at least two consecutive timesteps (See Section \ref{centers_of_mass}).\\

Points 1-8 describe the general image processing routine with the Wavetrack framework as tested on AIA 193-channel data. Nevertheless, in principle any other FITS or PNG data can be used as an input.  Available image processing techniques can be applied in different sequences at any of the stages (1-8). Joined object masks produced with the multi-scale method can be split into intensity levels for them to be selected manually according to the setup. Median filtering (not used in events from this paper) and posterization can be applied straight at the stages 1-2, a one-pass image segmentation technique can be used instead of clustering, etc.\\

During the late stages of the shock wave propagation, for some events the pixel values of a certain regions of the shock wave (usually in the nose) could be below sensitivity of the overall computational scheme (See Fig \ref{fig:wavetrack_pairs_110511}, bottom left panel). In that case, several non-connected objects would still be detected by our algorithm and they would appear on the resulting mask within the final recomposition routine (see Fig \ref{fig:fig_scheme} (Pathway A) and algorithm description in Sections \ref{sect_structures_object_criteria} and \ref{multiscale_image_filtering}). Depending on the input data these parts may be revealed after additional adjustment of the decomposition parameters for detection of more dim objects.\\

At the final stage we estimate the object velocity by calculating the speed of the center of mass of a CBF mask applied to raw data for at least two consecutive time steps. For each object mask a "center of mass" is calculated based on pixel coordinates with pixel intensities taken as weights. In binary object masks, all pixel weights are equal and calculating coordinates of the "center of mass" in this case would reveal the geometrical center of the object.

\subsubsection{Single Timestep Processing Stages and Parameters}
\label{processing_stages}
Here we use three timesteps of the June 07, 2011 AIA-observed CBF event to illustrate the various processing stages of the Wavetrack framework. 
\begin{figure}[htp]
    \centering
    \includegraphics[width=16cm]{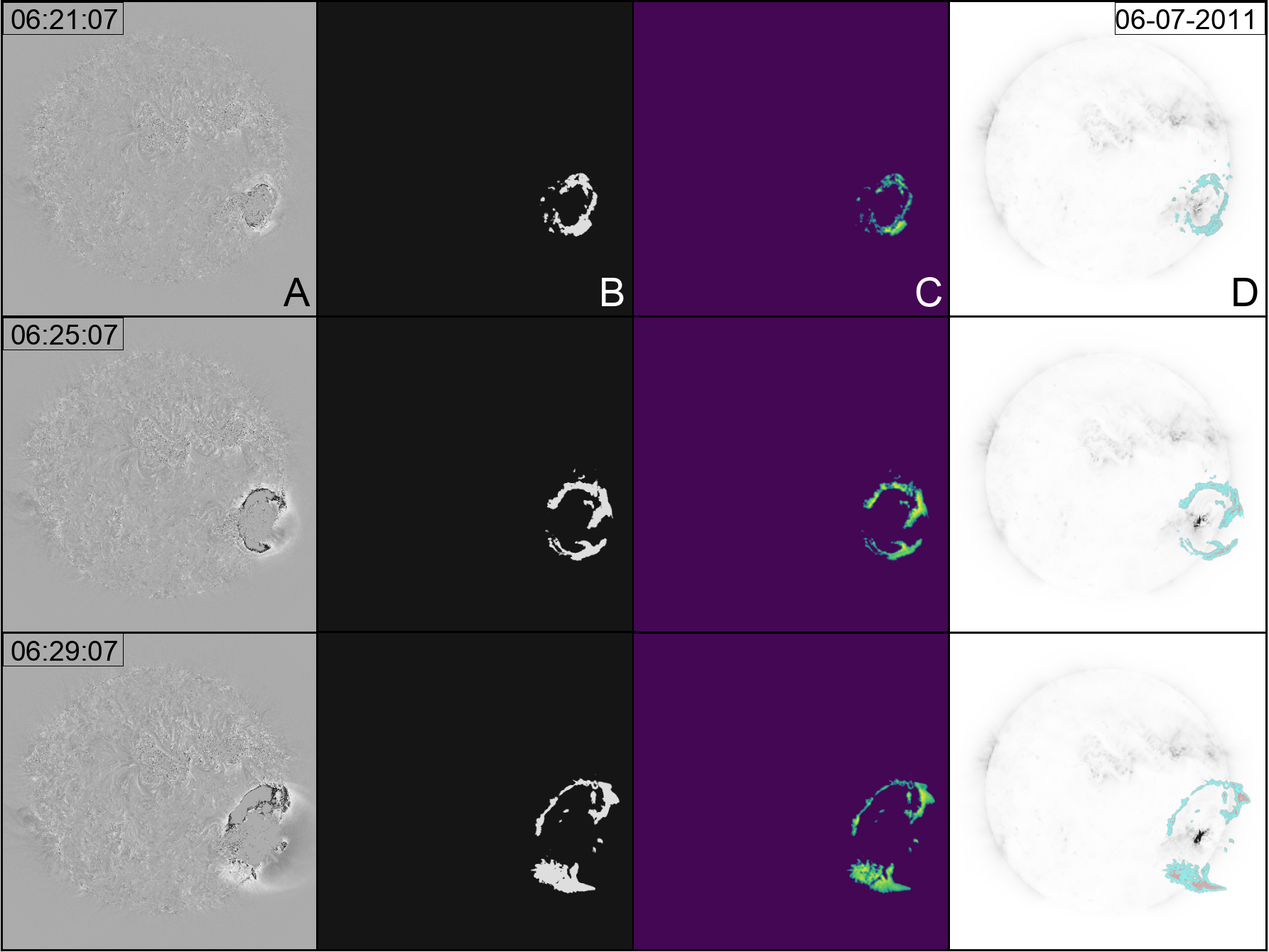}
    \caption{Demonstration of the Wavetrack application stages, for four times during the event of June 07, 2011. From left to right, the columns show: A) base difference images, B) recomposed object masks, C) object masks applied to the original data, and D) the applied masks overlaid on base difference images.}
    \label{fig:wavetrack_stages}
\end{figure}
Figure \ref{fig:wavetrack_stages}A shows base difference images obtained by subtracting a base image from each timestep image data and applying higher and lower thresholds for the absolute values of the pixels. 
A base image is created for each event as an average of a few consecutive timesteps 2-5 minutes prior to the beginning of the event. Absolute values of the threshold interval (-50,150) are selected for the purpose of narrowing of the image dynamic range and focusing on the image segment where the object of interest is located (i.e., the CBF in this case). 
At the next step (Fig. \ref{fig:wavetrack_stages}B), base difference images are decomposed with the \`a trous wavelet technique into a series of scales (see Fig. \ref{fig:a_trous}). To each of the wavelet coefficients a relative thresholding is applied once more depending on the statistical distribution of the pixel intensities for the each of decomposition levels. 
Finally, after segmentation the object masks at each time step are multiplied by the original data (Fig. \ref{fig:wavetrack_stages}C) to reveal the intensity of different parts of the object (CBF in this case). Fig. \ref{fig:wavetrack_stages}D shows the masks overlaid on the base difference images for context.

Here we described methods and algorithms implemented in our software, followed by single timestep example.  Exact processing and decomposition setup, and parameters of the filtering techniques depend on the input data and objects of interest (i.e – CME Shock Wave, Filament or an Active Region) and are given in examples section of the Wavetrack package \url{https://gitlab.com/iahelio/mosaiics/wavetrack}. 

\section{Application to Solar Eruption Observations}
\label{s3}

\subsection{Application to Coronal Bright Fronts}

We have applied the Wavetrack methodology to three previously studied eruptive events, observed by the AIA telescope \citep{Kozarev:2015,Kozarev:2017}. In the first event, which occurred on May 11, 2011, an erupting filament drives a relatively bright CBF on the northwest part of the solar disk close to the limb. The second event, on June 07, 2011, originated close to the southwest limb and featured a large and bright CBF. The third event occurred on December 12, 2013, and originated again close to the southwest limb.
\begin{figure}[htp]
    \centering
    \includegraphics[width=16cm]{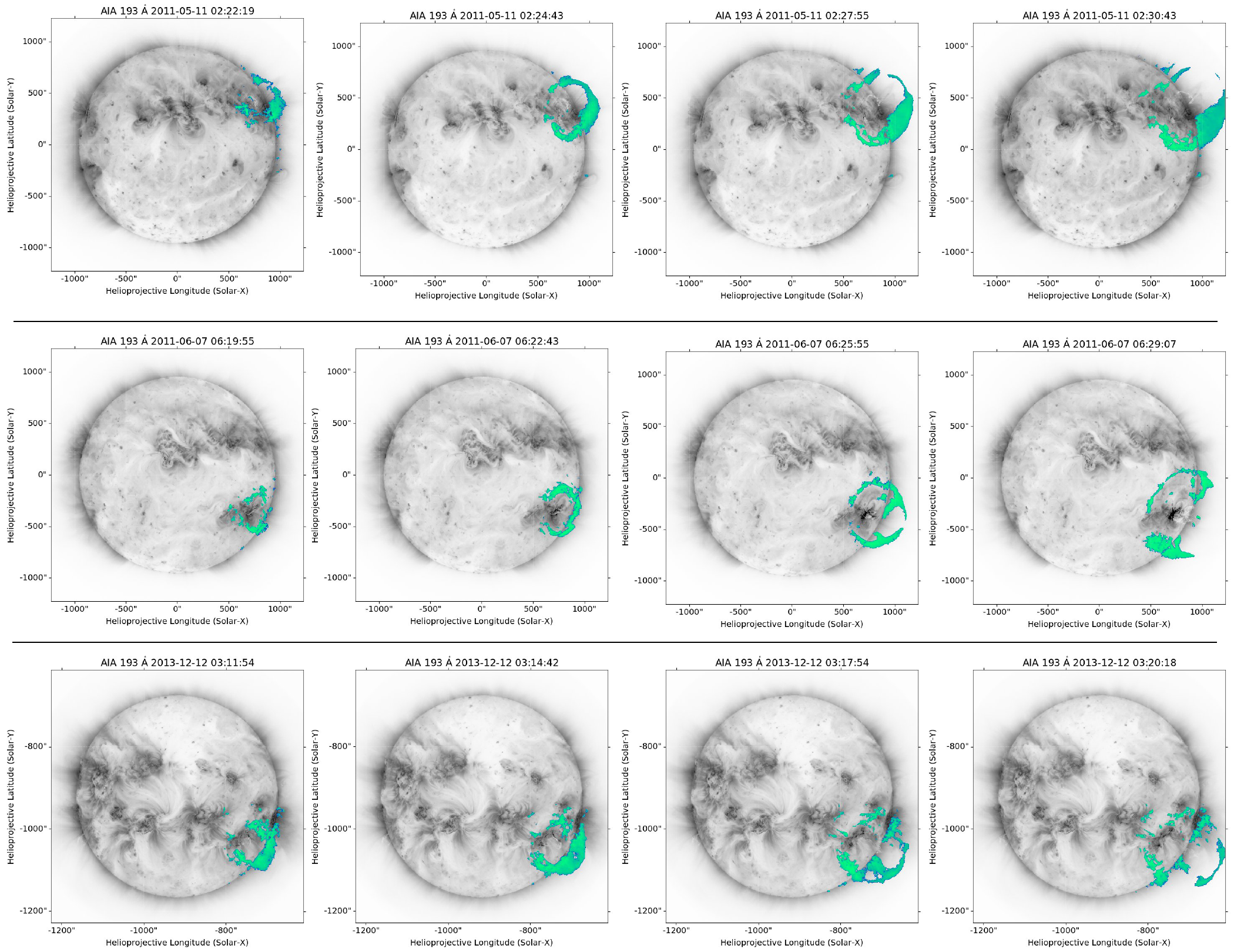}
    \caption{Output of the Wavetrack method for three separate CBF events: May 11, 2011; June 07, 2011; December 12, 2013. For each event, we show four snapshots of inverted-colormap grayscale AIA 193\AA~images, with the Wavetrack CBF masks applied to base difference data overlaid using a different colormap to highlight the them.}
    \label{fig:wavetrack_multievent}
\end{figure}
Figure \ref{fig:wavetrack_multievent} shows the successfully extracted Wavetrack CBF objects for the three events, for four observations separated by 3 minutes each. The Wavetrack object masks were applied to base difference images and visualised in a blue-green color map over the original AIA 193\AA~observations in grayscale colormap, to accentuate the CBFs. In all three events, Wavetrack is able to capture very well the extent of the CBF in the consecutive time steps. It easily follows the evolution of the CBFs both on the solar disk and off the limb, even though the pixel distributions and intensities in those two regions are very different. These applications allow us to study in detail the time-dependent shapes of the CBFs, as well as their changing intensity distributions, separate from the rest of the corona. In addition, the Wavetrack method allows to successfully select the CBF objects under different coronal activity states. During the December 12, 2013, the corona is much more active in this EUV channel than the previous two events, but this does not preclude the successful segmentation of the CBF. Finally, the segmentation of the CBF to the last event demonstrates that Wavetrack is capable of tracking multiple separate parts of the same feature.

\subsection{Application to Eruptive Filaments}
\begin{figure}[htp]
    \centering
    \includegraphics[width=16cm]{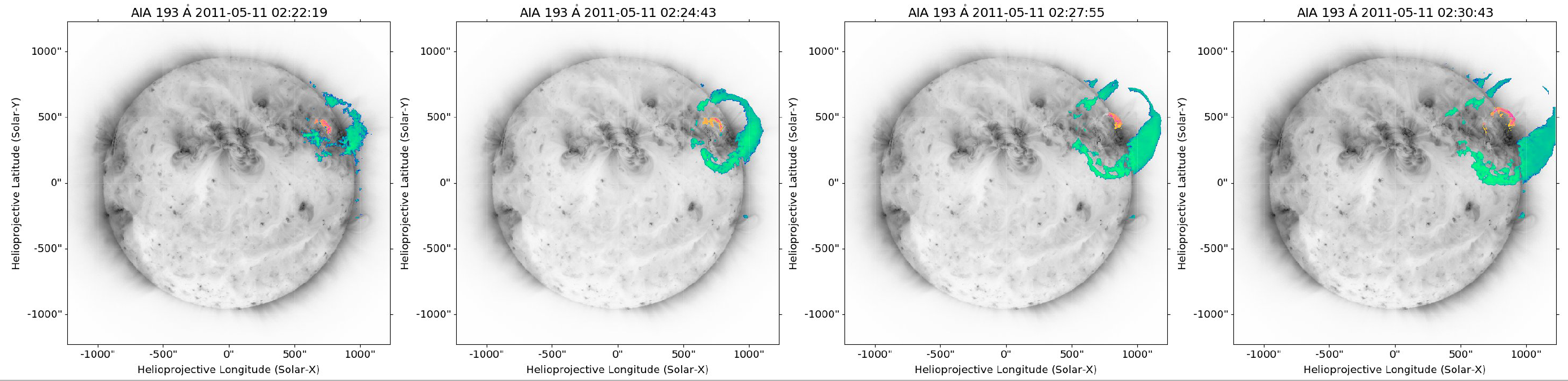}
    \caption{Combined tracking of the May 11, 2011 CBF and the filament driving it, with Wavetrack is shown here for three snapshots of the event, separated by approximately 2 minutes each.}
    \label{fig:wavetrack_cbf_filament}
\end{figure}

Wavetrack may be used to isolate different observational features in the same eruptive event, allowing to study the spatial and temporal relationships between them. This is accomplished by executing a separate run of the algorithm with appropriate input parameters for tracking filament material. Figure \ref{fig:wavetrack_cbf_filament} shows three time steps from the May 11, 2011 event, for which both the erupting filament (in orange-red) and the CBF it drives (in blue-green) have been segmented from AIA 193\AA~observations and overlaid on the original AIA data. This allows to follow the relationship between the driver and wave, demonstrating the versatility of Wavetrack.

\begin{figure}[htp]
    \centering
    \includegraphics[width=16cm]{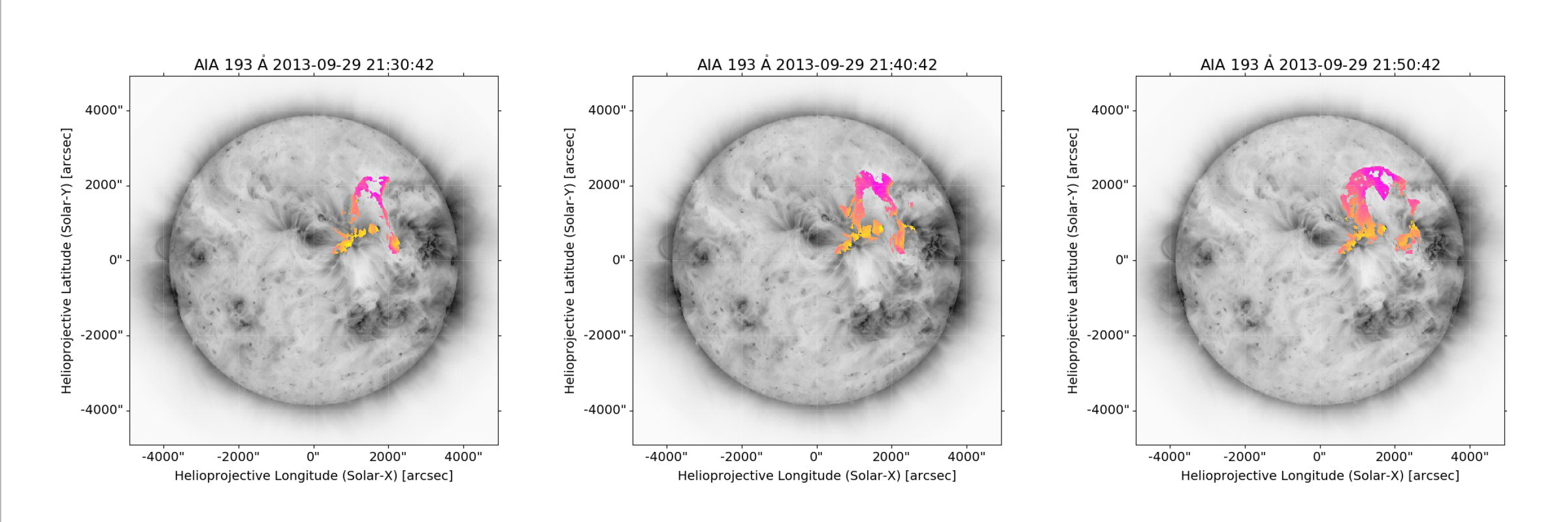}
    \caption{Tracking of the September 29, 2013 large-scale eruptive filament  with Wavetrack is shown here for three snapshots early in the event, separated by 10 minutes each.}
    \label{fig:wavetrack_filament}
\end{figure}
The presented applications of Wavetrack to solar eruptive events demonstrate the versatility of Wavetrack - it recognizes and tracks the desired features on the solar disk and off the solar limb. 
These features of Wavetrack make it perfect for inclusion in pipelines aimed at faster processing and targeted analysis with additional models such as differential emission measure (DEM) and Fourier local correlation tracking (FLCT). Below, we extend the analysis by examining the plane-of-sky velocities of the studied events.

\subsection{Estimation of Plane-of-Sky Velocities}

We further apply the Wavetrack code to isolating and studying the kinematics of CBFs and filaments in detail. To do that, we employ the Fourier Local Correlation Tracking method \citep{Welsch:2004, Fisher:2008}, used extensively for the estimation of horizontal flows in photospheric magnetograms. The method was initially proposed as a cross-correlation technique for the precise measurement of the proper motion of traces seen on successive images of time series of solar granulation \citep{November:1988}. The aim of FLCT calculation for two consecutive 2D maps of some scalar quality is to find a 2D flow field which, when applied to the scalar field in 1st image will most closely resemble the 2nd image. To construct a 2D velocity field that connects two images $I_1(x,y)$ and $I_2(x,y)$ taken at two different times $t_1$ and $t_2$, the algorithm starts from some given location within both images, the velocity vector is calculated, and then this step is repeated varying the location over all pixel positions. When applied to the object masks output by the Wavetrack code, it produces detailed maps of the instantaneous velocity of the object of interest. In this application, we use a freely available Python implementation of the method as a Sunpy-affiliated package \footnote{https://github.com/sunpy/pyflct/}.

\begin{figure}[htp]
    \centering
    \includegraphics[width=11cm]{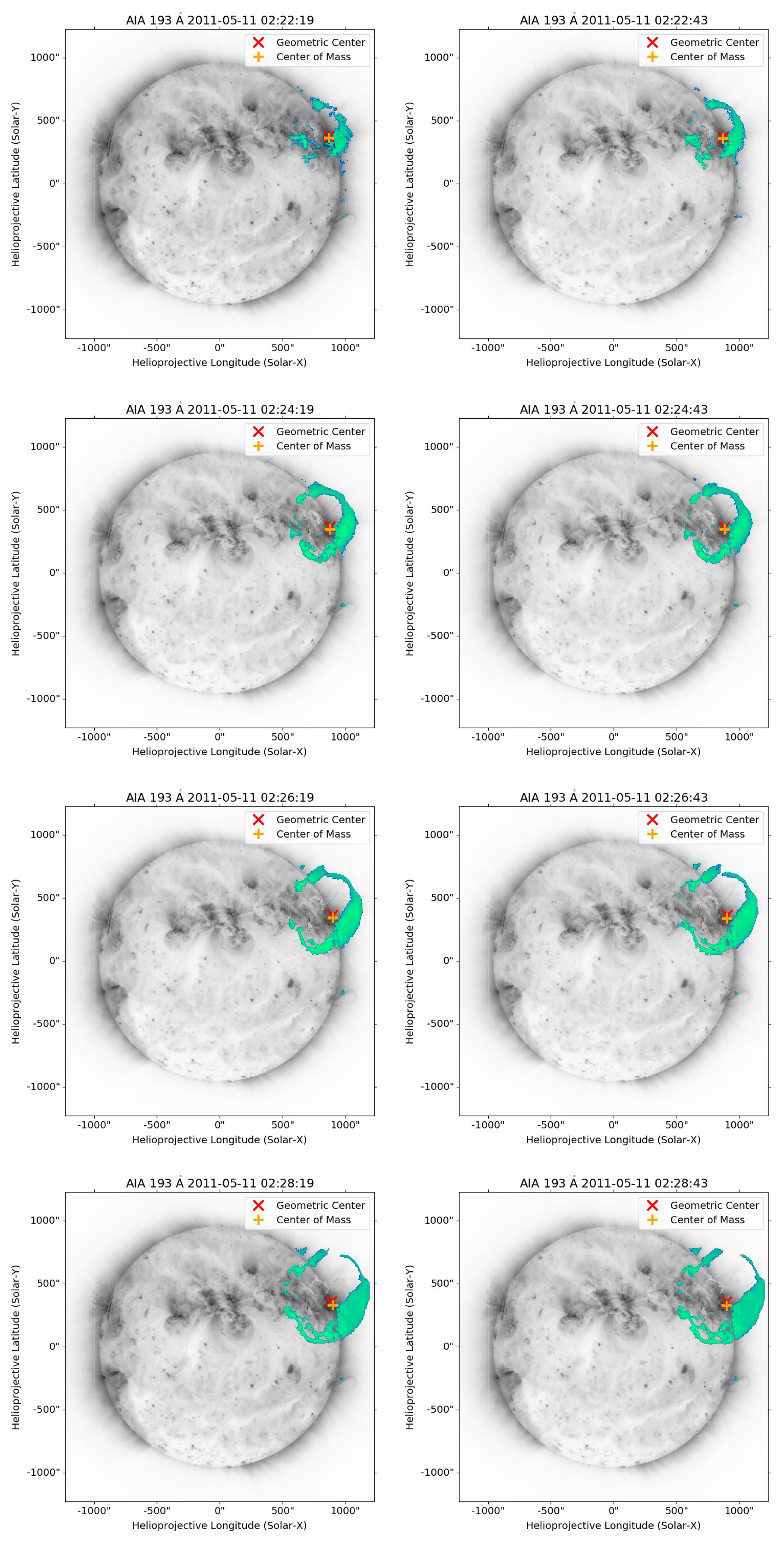}
    \caption{Wavetrack output shown for four pairs of consecutive images separated by two minutes, during the eruption of May 11, 2011, with the applied CBF mask shown similarly to Fig. \ref{fig:wavetrack_multievent}.}
    \label{fig:wavetrack_pairs_110511}
\end{figure}

Figure \ref{fig:wavetrack_pairs_110511} shows four pairs of consecutive AIA images (one pair per row) from the May 11, 2011 event, used as input for the FLCT algorithm. The images in each pair are separated by 24 sec. The pairs in the figure are two minutes apart from each other, in order to follow the CBF progression over 6 minutes. The CBF change is difficult to observe in each pair, but is significant between pairs. 

\begin{figure}[htp]
    \centering
    \includegraphics[width=12cm]{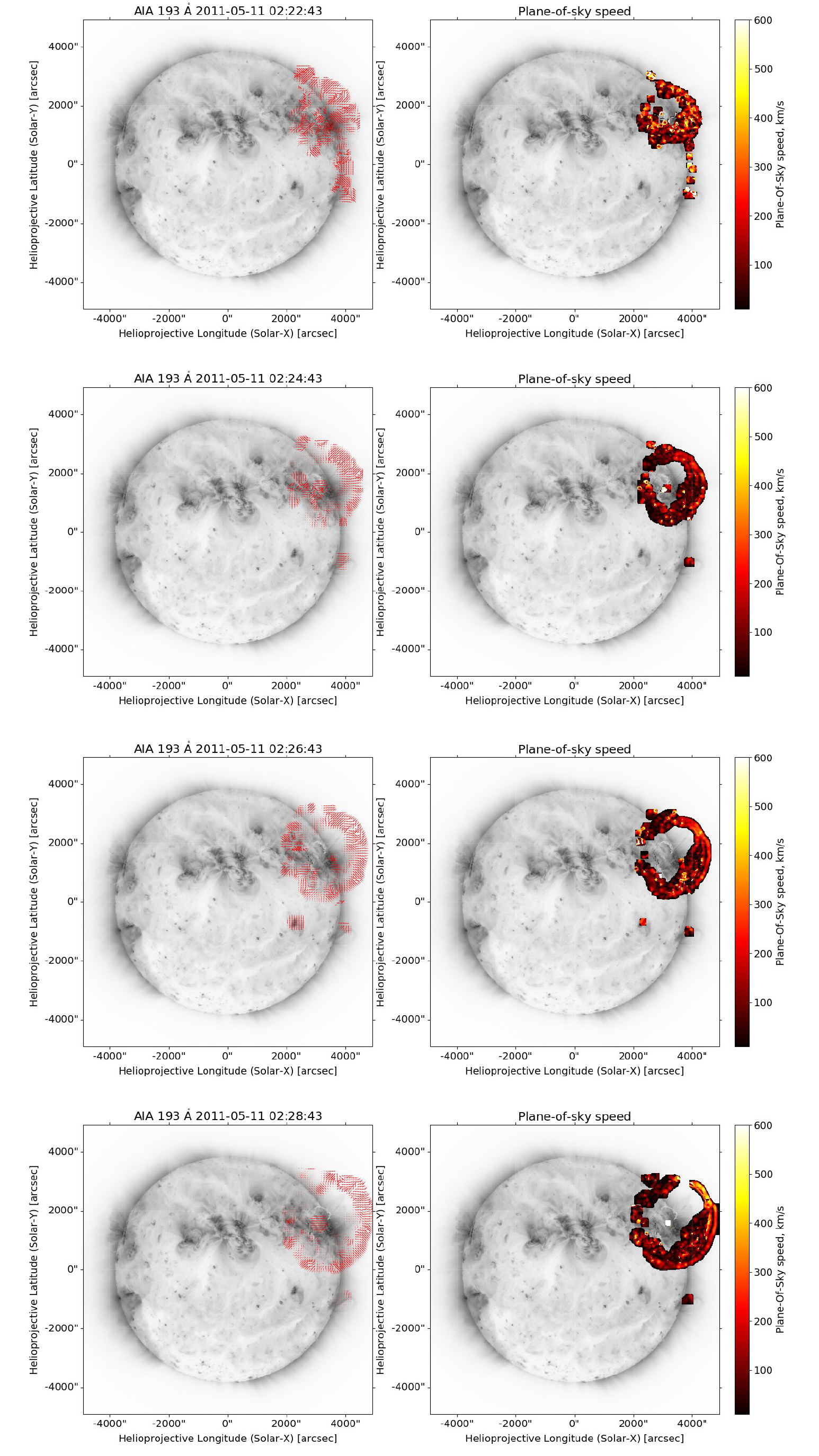}
    \caption{Output of the FLCT model for the four pairs of consecutive images shown in Fig. \ref{fig:wavetrack_pairs_110511}. Left: arrows showing the plane-of-sky velocity. Right: the plane-of-sky speed.}
    \label{fig:flct_110511}
\end{figure}

The corresponding FLCT results are shown in Fig. \ref{fig:flct_110511}. Each row corresponds to a row in Fig. \ref{fig:wavetrack_pairs_110511}. The left columns show the instantaneous plane-of-sky direction of motion of each region of the CBF, using red arrows. The arrows' lengths are scaled with the local speed. We have removed spurious vectors with speeds larger than 1500 km/s. In the right columns the instantaneous speed map is shown for the same four times. The color map is scaled between 0-600~km/s from black to white. The results show clearly how the CBF is expanding from the central source in an uneven fashion. What is more, in this event the thinnest part of the CBF corresponding to the region strongly driven by the erupting filament, is also the fastest moving as the event processes - in contrast with the other regions both on disk and off limb. This cannot be inferred from intensity observations alone, but it is easy to observe in these results.

\begin{figure}[htp]
    \centering
    \includegraphics[width=11cm]{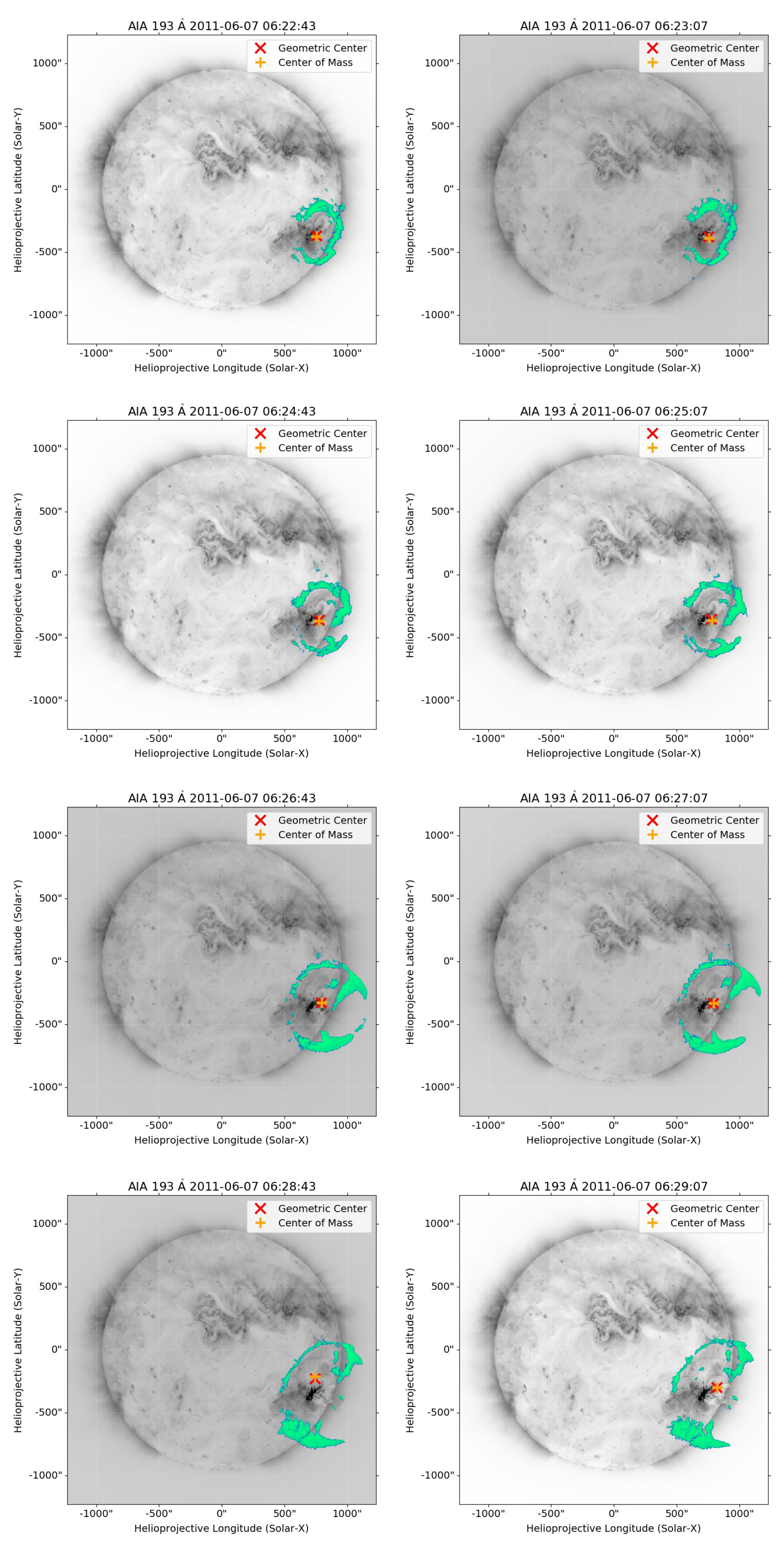}
    \caption{Wavetrack output shown for four pairs of consecutive images separated by two minutes, during the eruption of June 07, 2011, similar to Fig. \ref{fig:wavetrack_multievent}}
    \label{fig:wavetrack_pairs_110607}
\end{figure}

\begin{figure}[htp]
    \centering
    \includegraphics[width=12cm]{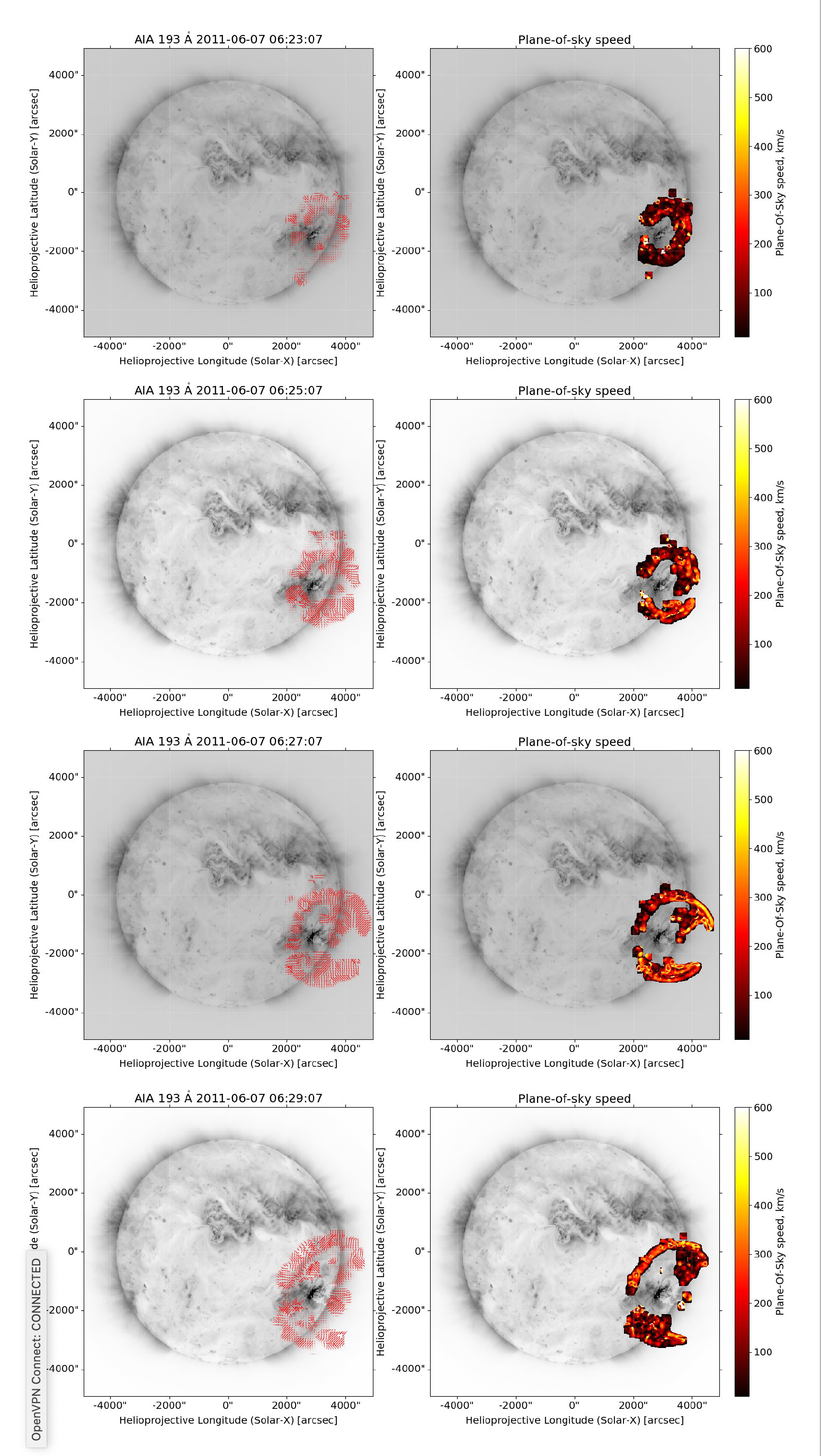}
    \caption{Output of the FLCT model for the June 07, 2011 event, for the four pairs of consecutive images shown in Fig. \ref{fig:wavetrack_pairs_110607}. Left: arrows showing the plane-of-sky velocity. Right: the plane-of-sky speed.}
    \label{fig:flct_110607}
\end{figure}

For the event on June 07, 2011, the Wavetrack pairs are shown in Figure \ref{fig:wavetrack_pairs_110607}. Again, every two consecutive images are separated by 24~sec, while the pairs are separated by 2~minutes. The FLCT results are shown in Figure \ref{fig:flct_110607}. Here, a clear increase in the average speed can be seen as the event progresses - first in the leading edges moving away from the Sun, and later in the on-disk front moving towards the center of the solar disk.

\begin{figure}[htp]
    \centering
    \includegraphics[width=11cm]{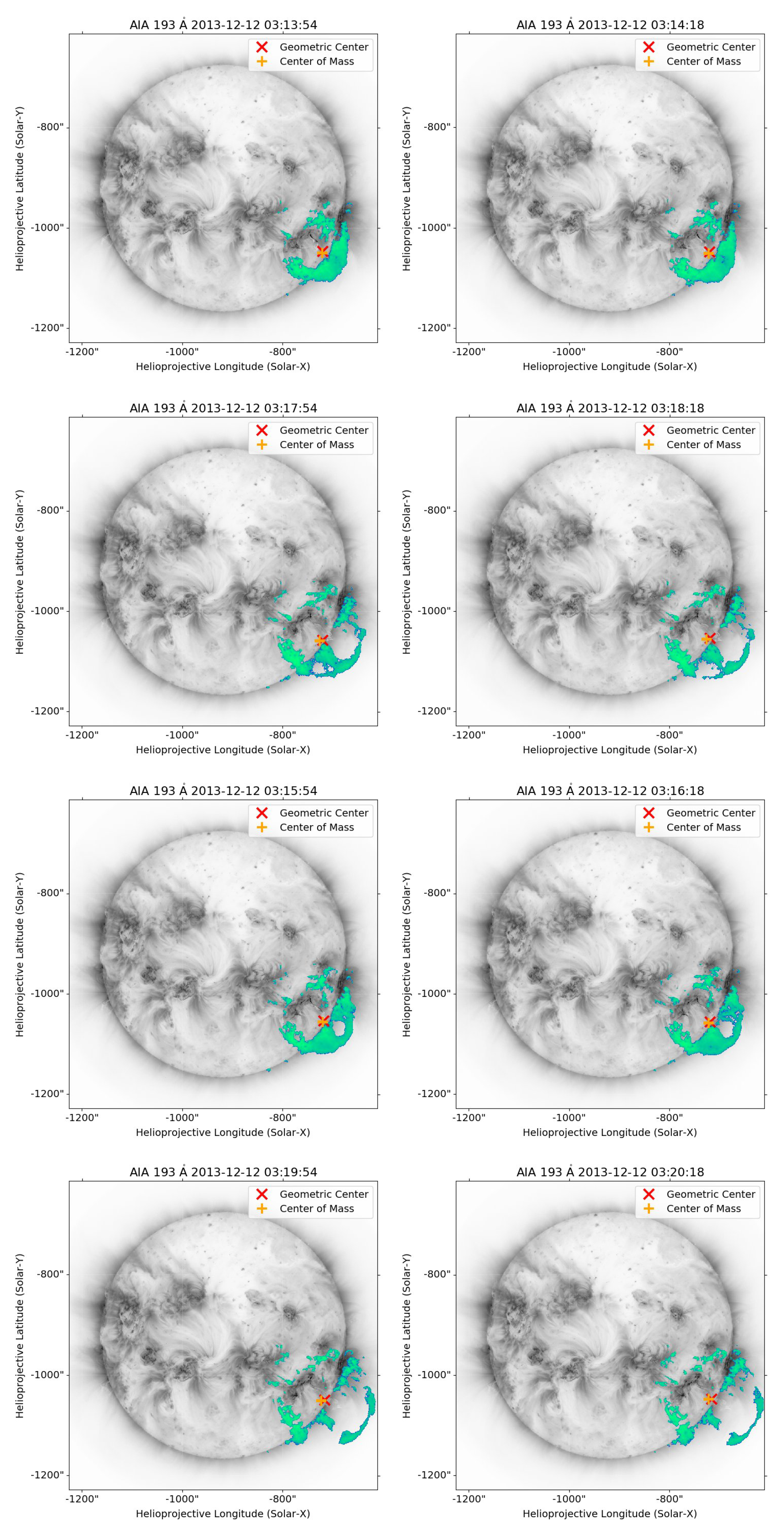}
    \caption{Wavetrack output shown for four pairs of consecutive images separated by two minutes, during the eruption of December 12, 2013, similar to Fig. \ref{fig:wavetrack_multievent}}
    \label{fig:wavetrack_pairs_131212}
\end{figure}

\begin{figure}[htp]
    \centering
    \includegraphics[width=12cm]{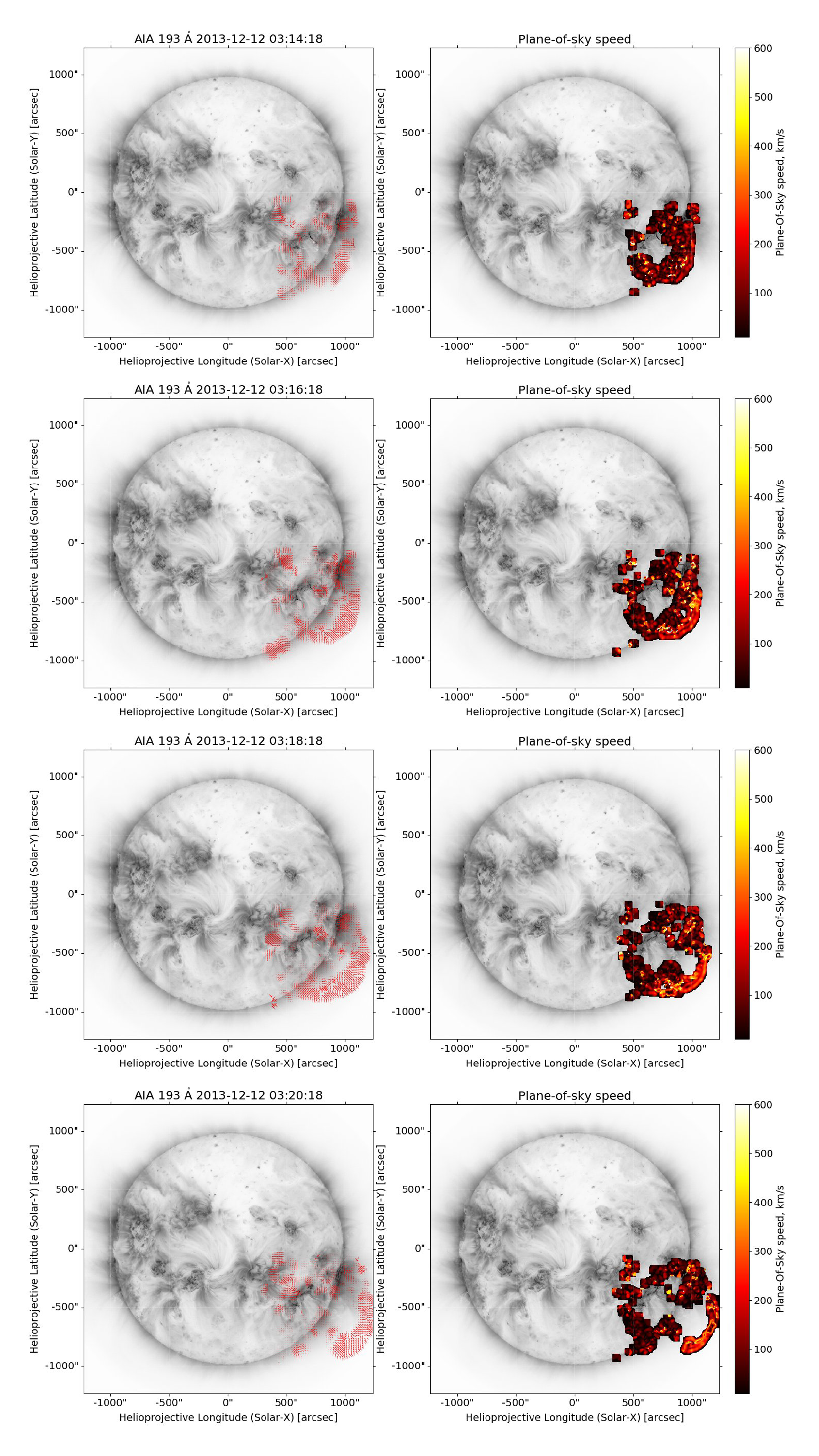}
    \caption{Output of the FLCT model for the December 12, 2013 event, for the four pairs of consecutive images shown in Fig. \ref{fig:wavetrack_pairs_131212}. Left: arrows showing the plane-of-sky velocity. Right: the plane-of-sky speed.}
    \label{fig:flct_131212}
\end{figure}

Finally, the Wavetrack pairs for the December 12, 2013 event are shown in Figure \ref{fig:wavetrack_pairs_131212}. They are again separated by 2~min each, while consecutive images are separated by 24~sec. Here the CBF quickly becomes split into separate sub-features, likely due to the interactions of its various parts with the much more dynamic underlying coronal environment. The active magnetic environment surrounding the eruption source is easily seen. This may be the cause of the slower speeds observed here, shown in Figure \ref{fig:flct_131212}. Similarly to the May 11, 2011 event, the highest speeds are observed consistently in the direction away from the Sun, above the erupting filament driver. Again, the continued compression by the driving filament has caused a thinner wave front region, as well as higher speeds there.

\begin{figure}[htp]
    \centering
    \includegraphics[width=12cm]{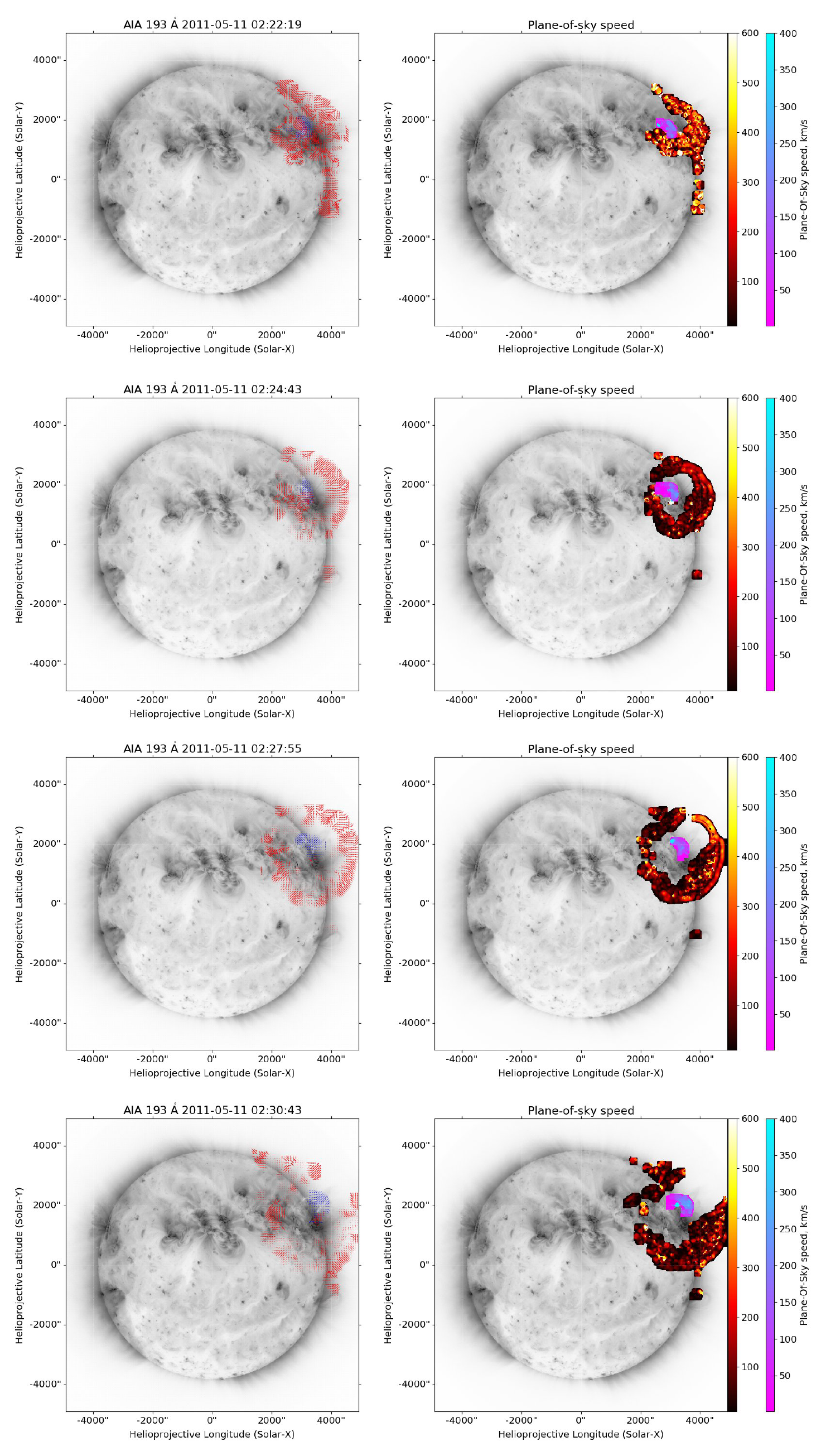}
    \caption{Output of the FLCT models for the combined CBF/filament May 11, 2011 event, for the three images shown in Fig. \ref{fig:wavetrack_cbf_filament}. Left: arrows showing the plane-of-sky velocity vectors (red color for the CBF object, blue for the filament). Right: the plane-of-sky speed.}
    \label{fig:flct_wave_filament_110511}
\end{figure}

We explore further this effect in the May 11, 2011 event, where we have also calculated the plane-of-sky velocities of the erupting filament driver. Figure \ref{fig:flct_wave_filament_110511} shows overlaid three time steps of the event with the velocity vectors on the left panels for the CBF (in red) and the filament (in blue); the right panels show the speed maps for the CBF as in Fig. \ref{fig:flct_110511} (in yellow-red-black color map), and the filament (in turquoise-blue-purple color map). Our method allows to demonstrate that the region of highest speeds in the rising filament (darker blue pixels in the right plots) is directly below the region of highest speeds in the CBF (brighter yellow pixels).

\subsection{Kinematics of CBF Centers of Mass / Geometric Centers}
\label{centers_of_mass}

Another aspect of the Wavetrack method that can be used is the determination of instantaneous positions and speeds of the feature centers. The method provides as output the geometric centers (GC) and the centers of mass (CM) for the three events studied here. The GC is defined as the geometric center of all Wavetrack mask pixels at a given time, while the CM also includes a weighting based on the AIA 193\AA~channel's base difference intensity. The GC may be used to determine whether the CBF center is moving, which is useful for geometric modeling of compressive fronts in events. On the other hand, the instantaneous difference between the GC and CM positions is due to an activation manifested by increase in the intensity or emitting area of certain regions of the CBF. This may further be explored by calculating the angle between these two points over time, to provide a measure of the relative activity as a function of position along the CBF.

\begin{figure}[htp]
    \centering
    \includegraphics[width=10cm]{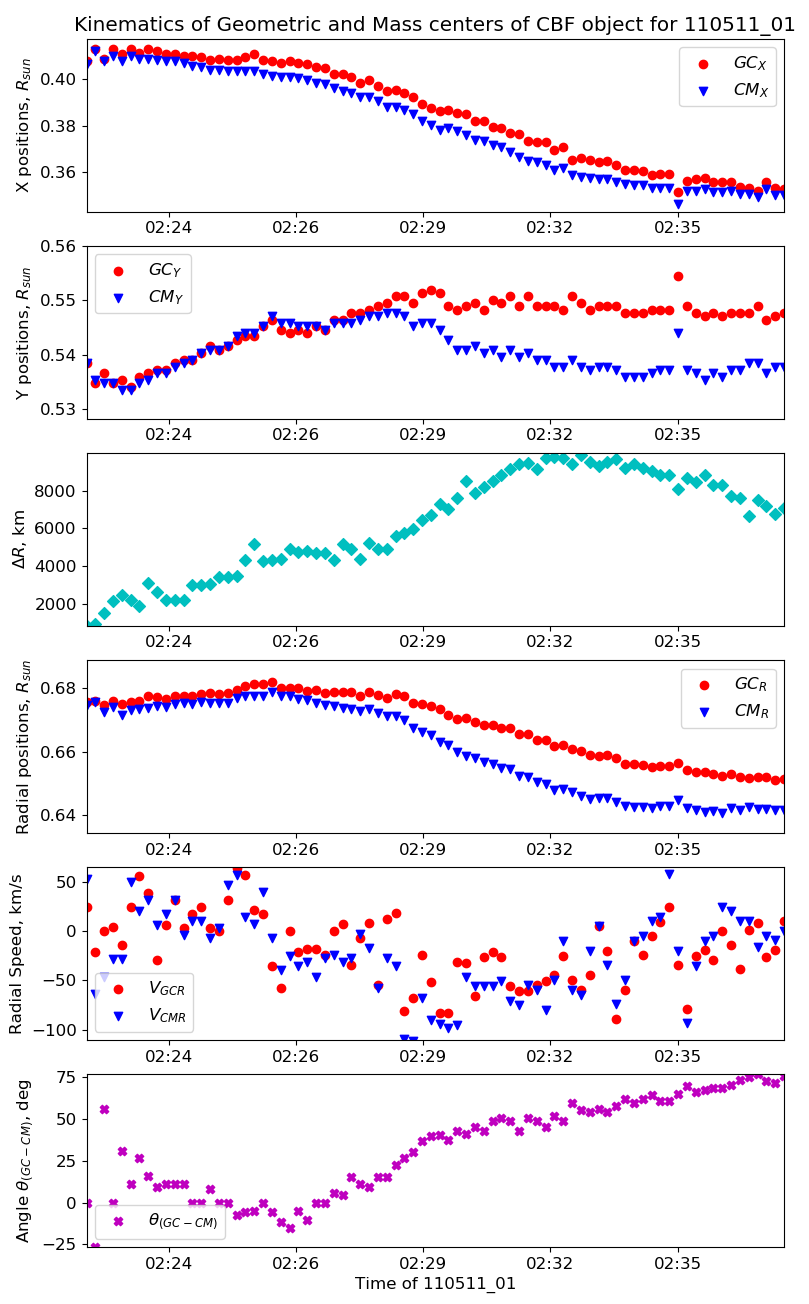}
    \caption{Kinematics of the center of mass and geometric centers for the event on May 11, 2011. The rows show GC and CM X-, Y-, and R- positions, the magnitude of the distance between the GC and CM in km, and the angle between the two points as a function of time.}
    \label{fig:centers_of_mass_110511}
\end{figure}

We provide, for the three events studied here, $GC$ and $CM$ metrics over time. Figure \ref {fig:centers_of_mass_110511} shows the results for the May 11, 2011 event. The panels show, from top to bottom: the X-axis ($GC_X$ and $CM_X$) and Y-axis ($GC_Y$ and $CM_Y$) positions, measured from the projected solar center in solar radii ($R_{sun}$); the radial distances $GC_R$ and $CM_R$, in $R_{sun}$; the distance between the two radial positions $\delta r=GC_R-CM_R$, in km; the radial speeds of the two points, $V_{GCR}$ and $V_{CMR}$, in km/s; and the angle between the two position vectors $GC$ and $CM$ over time, $\theta_{GC-CM}$, in degrees. A clear split between the two radial positions appears and increases after 02:26~UT. This is mostly due to a shift to the south of the CM. The evolution of the angle $\theta_{GC-CM}$ supports this change. The reason for this is the gradual increase in the thickness of the wave in the southwest of the front. At the same time, the speed of the GC and CM points varies only slightly, from 0 to 100 km/s.

\begin{figure}[htp]
    \centering
    \includegraphics[width=10cm]{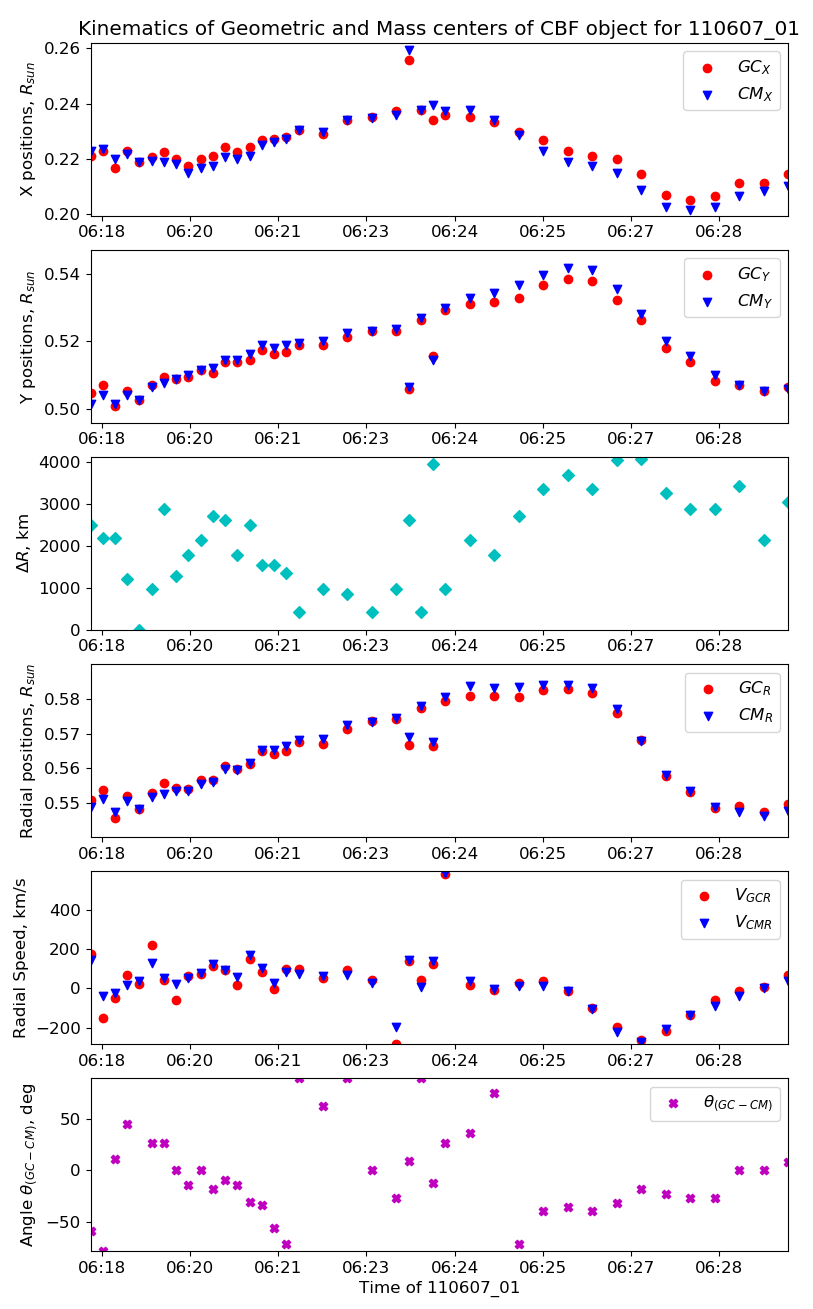}
    \caption{Kinematics of the center of mass and geometric centers for the June 07, 2011 event. The panels are the same as in Fig. \ref{fig:centers_of_mass_110511}.}
    \label{fig:centers_of_mass_110607}
\end{figure}

For the dynamic event on June 07, 2011, the situation is different, as can be seen from Figure \ref{fig:centers_of_mass_110607}. In that event, the GC and CM moved with speeds of up to 300~km/s in the radial direction. The angle between them changed quite a lot as different parts of the CBF brightened and dimmed. We note that in this case, as well as with the following event, the significant decrease of the projected radial position of the GC and CM is mostly due to the north-west parts of the CBF dimming and disappearing from the Wavetrack feature. However, in the same period the angle $\theta$ between the two remains robust, changing only slightly.

\begin{figure}[htp]
    \centering
    \includegraphics[width=10cm]{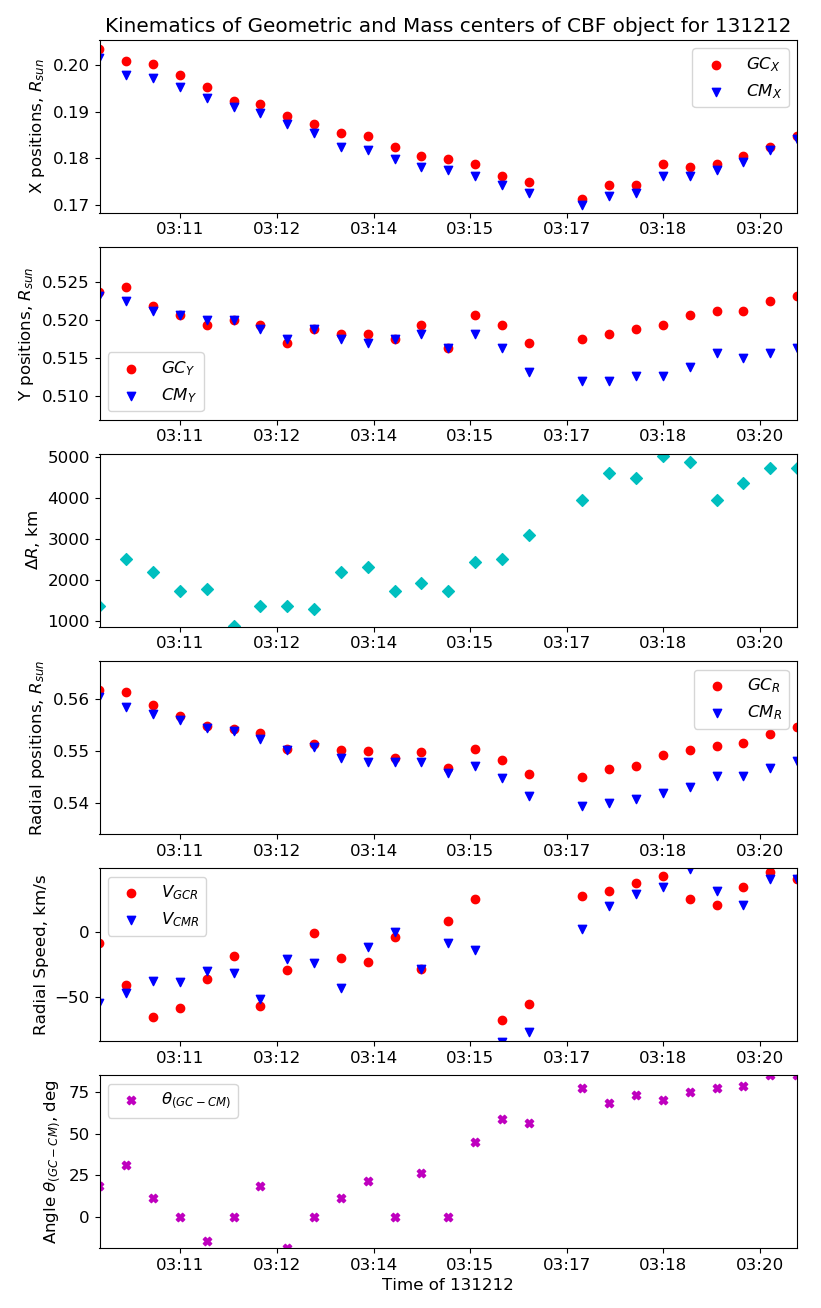}
    \caption{Kinematics of the center of mass and geometric centers for the December 12, 2013 event. The panels are the same as in Fig. \ref{fig:centers_of_mass_110511}.}
    \label{fig:centers_of_mass_131212}
\end{figure}

Finally, in the event on December 12, 2013, we observe smoothly varying positions of the GC and CM, and relatively slow speeds under 100~km/s. The increase of the radial position after 03:17~UT we attribute to the decrease in brightness of the southern part of the CBF. The angle $\theta$ is decreasing for most of the event, pointing increasingly southward, in line with the observed feature evolution.

\section{Summary and Conclusions}
\label{s4}

In this work we have presented Wavetrack, a new method for the automated detection and tracking of dynamic coronal features. The method utilizes wavelet decomposition, feature enhancement and filtering, and final object recomposition. Wavetrack produces time-dependent feature pixel masks, which can be applied to integral or base-difference images to produce final feature maps. It is capable of tracking the pixels related to very dim, large-scale features like coronal bright fronts / EUV waves in AIA observations. It has also been applied successfully to eruptive filaments in these data. Wavetrack works for both on-disk and off-limb features, and it is capable of tracking well features that split into separate parts over time. We have implemented it as a flexible, object-oriented framework written in Python, and it is freely available for download and use.

We have applied Wavetrack to four separate events, of which we focus on three -- the CBFs on May 11 and June 07, 2011; and on December 12, 2013. For all three, the method tracks the full CBF pixel maps. The model results have been used as input to the FLCT method of calculating plane-of-sky speeds. Combined with the Wavetrack results, its application readily reveals the evolution of driven and non-driven regions of the CBFs, as well as their relation to eruptive filament drivers. We find that drivers cause a compression effect thinning the CBF and increasing its speed, as expected from models. However, CBF brightness does not necessarily increase significantly in the driven regions in AIA data.

Finally, the method allows to track the changes over time as a function of the feature regions, by calculating the time-dependent vector between the pixel geometric center and center of mass, computed by weighing in the observed pixel intensities. This is especially useful for large-scale features such as CBFs, as it provides a simple metric with a one-dimensional time series for characterizing the feature evolution.

Currently, the model is somewhat limited in its application by the necessity of human input, in order to be used for segmentation of specific types of features, especially dim ones. Several object criteria, such as the expected threshold intervals and the weight coefficients of the recomposition, must be set manually. When base difference imaging is used, the base image must be chosen relatively precisely as not to pollute the input data with spurious features. Finally, in some cases it is necessary to fine-tune parameters for a specific event by a trial and error method. These limitations will be addressed in future work.

Our method is widely applicable both to different types of solar dynamic features, and to different observational data. In future work, we intend to apply it to detailed studies of filament evolution, as well as to coronagraph data, with the goal of improving our understanding of how eruptive large-scale fronts change between different types of observations, as well as between the low and middle corona.

\appendix
\section{Multi-Scale Data Representation Approach and a Wavelet Transform}
Generally, the multi-scale image analysis/decomposition concept is based on the idea that every Hilbert space $L^{2}(R)$ could be presented through hierarchy of orthogonal sub-spaces $V_{m}$ while in closure, one can assume that $L^{2}(R)$ can be represented in the following way:  

\begin{equation}
V_{2} \subset V_{1} \subset V_{0} \subset V_{-1} \subset V_{-2} \subset \ldots, \bigcap_{m \in Z} V_{m}=\{0\}, \bigcup_{m \in Z}=L^{2}(R)
\label{eq:a1}
\end{equation}

Here, for simplicity we considered the case of 1d signal. For every function which belongs to $V_{m}$ there is a corresponding scaled function so that:

\begin{equation}
f(x) \in V_{m} \longleftrightarrow f(2 x) \in V_{m-1}
\label{eq:a2}
\end{equation}

and there exists a function

\begin{equation}
\phi_{0}(x) \in V_{0}
\label{eq:a3}
\end{equation}

so that its negative increment

\begin{equation}
\phi_{0, n}(x)=\phi(x-n) \mathrm{n} \in \mathrm{Z}
\label{eq:a4}
\end{equation}

would form an orthogonal basis of the space, which means that functions

\begin{equation}
\phi_{m, n}(x)=2^{-m / 2}, \phi\left(2^{-m} x-n\right), \mathrm{n} \in \mathrm{Z}
\label{eq:a5}
\end{equation}

form an orthogonal basis of the space $V_{m}$ 
Every function $\mathrm{f}(\mathrm{x}) \in L^{2}(R)$ could be presented as a set of approximations $f_{m}(x) \in V_{w}$, i.e.

\begin{equation}
f(x)=\lim _{x \rightarrow \infty} f_{m}(x)
\label{eq:a6}
\end{equation}

Thus, the function $f_{m}(x)$ is an orthogonal projection of function f(x) on space $V_{m}$

\begin{equation}
f(x)-\sum_{n}\left\langle\phi_{m, u}, f(x)\right\rangle \phi_{m, n}(x)-\sum_{n} c_{m, n} \dot{\phi}_{m, n}(x)
\label{eq:a7}
\end{equation}

as,

\begin{equation}
\phi(x)-\phi_{0,0}(x) \in V_{0} \subset V_{-1}
\label{eq:a8}
\end{equation}

and scaling equation is written as

\begin{equation}
\phi_{0.0}(x)-\sqrt{2}  \sum_{n} h_{n} \phi_{-1, n}(x)-2 \sum_{n} h_{n} \psi(2 x-n)
\label{eq:a9}
\end{equation}

Now, define another object, wavelet spaces $W$, so that $L^{2}(R)$ can be decomposed as 

\begin{equation}
L^{2}(\mathbb{R})=\sum_{j \in \mathbb{Z}}^{\bullet} W_{j}:=\cdots \dot{+} W_{-1}+W_{0} \dot{+} W_{1} \dot{+} \cdots
\label{eq:a10}
\end{equation}

where every function f(x) which belongs to $L^{2}(R)$ has its unique decomposition 

\begin{equation}
f(t)=\cdots+g_{-1}(t)+g_{0}(t)+g_{1}(t)+\cdots
\label{eq:a11}
\end{equation}

where $g_{j} \in W_{j}$.
A wavelet function that satisfies the scaling condition \ref{eq:a9} is

\begin{equation}
\frac{1}{2}\left(\frac{x}{2}\right)=\sum_{i} g(k) \phi(x-k)
\label{eq:a12}
\end{equation}

And wavelet coefficients are calculated as
\begin{equation*}
w_{\jmath+1, l}=\sum_{k} g(k) c_{j, i+2^{i} k}
\end{equation*}

Each wavelet decomposition level could be represented by subtracting the data on j and j+1 levels:

\begin{equation}
w_{j+1}=c_{i 1}-c_{j+1,l} 
\label{eq:a13}
\end{equation}

The difference $C_{0,1}-C_{1,1}$ represents difference in scales, for wavelet decomposition $\varphi(\mathrm{x})$, while corresponding wavelet function is defined as

\begin{equation}
\frac{1}{2} \psi\left(\frac{x}{2}\right)=\phi(x)-\frac{1}{2} \phi\left(\frac{x}{2}\right)
\label{eq:a14}
\end{equation}

The 1-d signal is obtained the following way:

\begin{equation}
c_{j, l}=\sum_{k} h(k) c_{j-1, l+2^{j k}}
\label{eq:a15}
\end{equation}

with $\{\mathrm{h}(\mathrm{k})\}$ calculated from the scaling function $\varphi(\mathrm{x})$

\begin{equation}
\frac{1}{2} \phi\left(\frac{x}{2}\right)=\sum_{l} h(l) \phi(x-l)
\label{eq:a16}
\end{equation}

In case of the \`a trous wavelet, f(x) takes form of a $B_{3}$ spline

\begin{equation}
B=\phi(x)-B_{3}(x)-\frac{1}{12}\left(|x-2|^{3}-4|x-1|^{3}+6|x|^{3}-4|x+1|^{3}+|x-2|^{3}\right)
\label{eq:a17}
\end{equation}

\section{Image Gradient Methods}
\label{image_gradient_appendix}

The gradient field of the image is the result of applying the gradient operator to an image. In this case, the gradient field of the final object image (after the object mask is applied to the original image) is used for visual interpretation but can also serve as additional criteria for more robust object segmentation, as it highlights object contours, allows to determine saddle point locations, and to calculate gradient vectors. Image gradients can also be used as part of training sets for deep convolutional networks. By default, we use the Sobel-Feldman operator, a discrete differentiation operator, to compute an approximation of the gradient of the image intensity function.  At each pixel of the image, the result would be either the corresponding gradient vector, or the norm of this vector. The magnitude 
\begin{equation}\label{eq:12}
\mathbf{G}=\sqrt{\mathbf{G}_{x}{ }^{2}+\mathbf{G}_{y}{ }^{2}}
\end{equation}
and direction 
\begin{equation}\label{eq:13}
\boldsymbol{\Theta}=\operatorname{atan}\left(\frac{\mathbf{G}_{y}}{\mathbf{G}_{x}}\right)
\end{equation}
of the vector are obtained as a product of separate vertical and horizontal convolutions of the image $\mathbf{A}$ with 3$\times$3 kernel matrices:
\begin{equation}\label{eq:14}
\mathbf{G}_{x}=\left[\begin{array}{lll}
+1 & 0 & -1 \\
+2 & 0 & -2 \\
+1 & 0 & -1
\end{array}\right] * \mathbf{A}
\end{equation}

\begin{equation}\label{eq:15}
\mathbf{G}_{y}=\left[\begin{array}{ccc}
+1 & +2 & +1 \\
0 & 0 & 0 \\
-1 & -2 & -1
\end{array}\right] * \mathbf{A}
\end{equation}
As the result of the operator is the product of two separate vertical and horizontal operations, it is relatively inexpensive in terms of computation and can be implemented in parallel threads. As an alternative, for less complex images the gradient is calculated by simple convolution of the image with the matrix
\begin{equation}\label{eq:16}
\mathbf{G}_{x,y}=\left[\begin{array}{ccc}
1 & 0 & -1 \\
0 & 0 & 0 \\
-1 & 0 & 1
\end{array}\right]* \mathbf{A}
\end{equation}
The curvature $\chi$ of the intensity at pixel coordinates (x, y) \citep{Hagenaar:1999} in a certain direction ($h_{i}$ , $h_{j}$) is given by 
\begin{equation}\label{eq:17}
\chi=2 * f(x, y)-f\left(x+h_{1}, x+h_{2}\right)-f\left(x-h_{1}, x-h_{2}\right)
\end{equation}
where
\begin{equation}\label{eq:18}
\left(h_{1}, h_{2}\right)=(1,0)(1,1)(0,1)(-1,1)
\end{equation}
Around a local maximum, the curvature must be positive in all directions. Initially, this approach was used for analysis of fluxes of concentrations on magnetograms. 
A similar routine was also implemented into the Wavetrack package and showed its efficiency on determining contours on AIA EUV images, as the gradient approximation given by the Sobel-Feldman operator is relatively rough, in particular for high-frequency variations in the image.

\section{Image Segmentation}
\label{image_segmentation_appendix}

Image segmentation is the classification of an image into different groups. In the current study it is done in two passes with a sliding window. At each iteration, non-zero intensity pixels are considered belonging to a new object with separately calculated gradient field value for the same coordinates used as additional criteria. Pixels considered belonging to an object are assigned with a new number. Eventually each object mask is placed on its own separate image. Ideally, if configuration parameters are precise enough, there would be only one object mask in the output. It would correspond to the shock wave, filament, etc. The rest of the pixel groups would form objects which are either too small, or would not pass previous filtering or decomposition-recomposition stages. In case of a somewhat less perfect setup, manual verification and selection of relevant objects from the output would be necessary.

Initial estimation of the objects count and locations is done via a Monte-Carlo algorithm with the window size $r$,
\begin{equation}\label{eq:19}
r=\sqrt{2 N_{\min }}
\end{equation}
where
$N_{\min}$
is the minimum number of pixels in an object. Pixels of the posterized image (see definition of posterization in  Section \ref{processing_stages}) are selected randomly, each at least $r$ pixels apart from the previous selection. Pixels with different intensity values and zero gradient field values are accepted as belonging to separate objects \citep{Metropolis:1949}.

\begin{acknowledgements}
This work is part of the MOSAIICS project, funded under contract \#KP-06-DV-8/18.12.2019 to the Institute of Astronomy and NAO, BAS, under the National Scientific Programme "VIHREN" in Bulgaria.
\end{acknowledgements}

\end{document}